\documentclass[useAMS,usenatbib]{mn2e}
\usepackage{graphicx}

\title[Near-IR photometry and spectroscopy of NGC~6539 and UKS~1]{Near-IR photometry and 
spectroscopy of NGC~6539 and UKS~1:    
two intermediate metallicity Bulge Globular Clusters
\thanks{Photometric data have been taken at the ESO/NTT Telescope,
        within the observing programme 73.D-0313.
	Spectroscopic data were obtained at the W.M.Keck Observatory,
        which is operated as a scientific partnership among the California
        Institute of Technology, the University of California, and the National
        Aeronautics and Space Administration. The Observatory was made possible
        by the generous financial support of the W.M. Keck Foundation.}}

\author[Origlia et al.]{L. Origlia$^1$, E. Valenti$^{2,1}$,
        R.~M. Rich$^3$, F.R. Ferraro$^2$  \\
 $^1$ INAF-Osservatorio Astronomico di Bologna, Via Ranzani 1, I-40127 Bologna,
      Italy, e-mail livia.origlia@bo.astro.it \\
 $^2$ Dipartimento Astronomia, Universit\`a di Bologna,  
      Via Ranzani 1, I-40127 Bologna, Italy, \\
      e-mail elena.valenti3@unibo.it,francesco.ferraro3@unibo.it \\
      $^3$ Department of Physics and Astronomy, University of California
      at Los Angeles, Los Angeles, CA 90095-1562, e-mail rmr@astro.ucla.edu \\
       }
\date{\today}

\begin{document}
\pagerange{\pageref{firstpage}--\pageref{lastpage}} \pubyear{2005}
\maketitle
\label{firstpage}

\begin{abstract}
Using the SOFI imager at ESO/NTT  
and the NIRSPEC spectrograph at Keck II, we have obtained
J,K images and echelle spectra covering the range $1.5-1.8~\mu \rm m$ for
the intermediate metallicity bulge globular clusters NGC~6539 and
UKS~1. 
We find [Fe/H]=--0.76 and --0.78, respectively,
and an average $\alpha$-enhancement of $\approx+0.44$~dex and 
$\approx+0.31$~dex,
consistent with previous measurements of metal rich bulge clusters, 
and favoring the scenario of rapid chemical enrichment.
We also measure very low  $\rm ^{12}C/^{13}C\approx 4.5\pm 1$ isotopic ratios 
in both clusters, 
suggesting that extra-mixing mechanisms due to {\it cool bottom processing}
are at work during evolution along the Red Giant Branch.
Finally, we measure accurate radial velocities of 
$\rm<v_r>=+31\pm 4~km/s$ and 
$\rm<v_r>=+57\pm 6~km/s$ 
and velocity dispersion of $\approx$8~km/s
and $\approx$11~km/s for NGC~6539 and UKS~1, respectively, 

\end{abstract}

\begin{figure*}
\centering
\includegraphics[width=18cm]{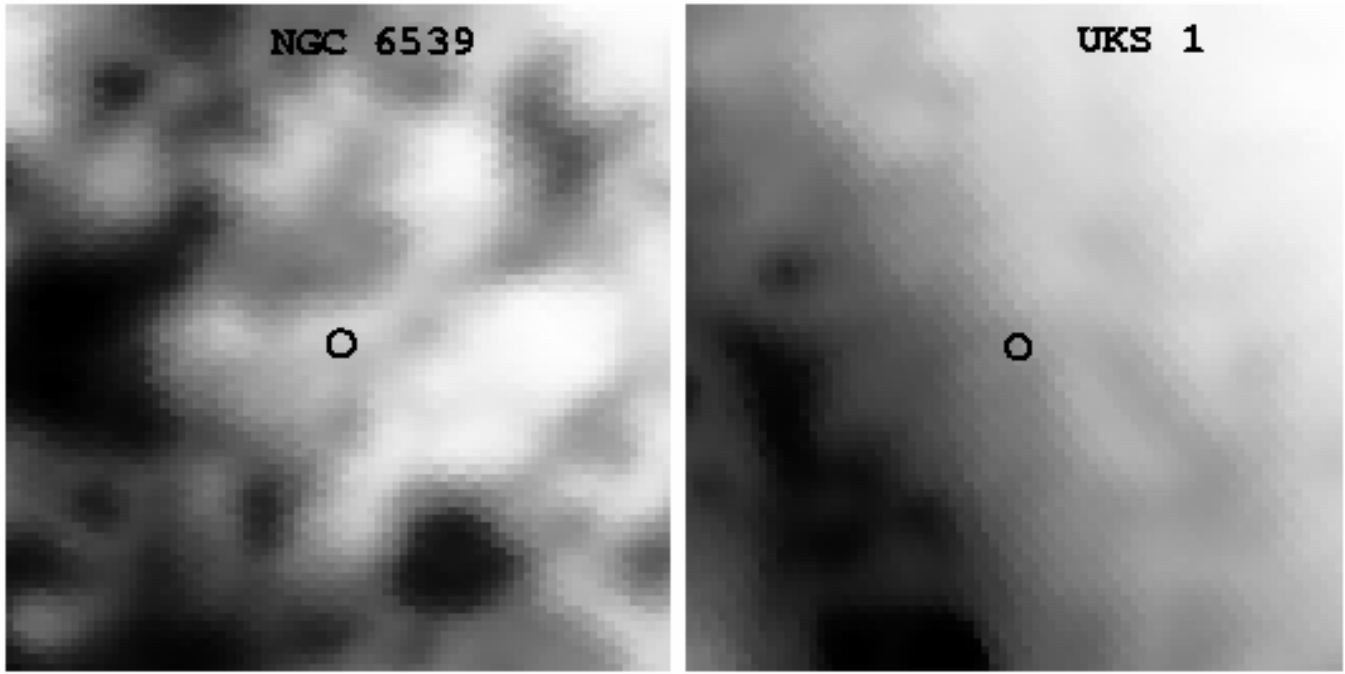}
\caption{
IRAS 100 $\mu$m images of $2^{\circ} \times 2^{\circ}$ fields around 
the clusters NGC~6539 (left) and UKS~1 (right), with 
the 100 $\mu$m emission (corresponding to the reddening) increasing 
from white to black. The empty circle marks the cluster position.
}
\label{fig1}
\end{figure*}

 
\section{Introduction}

Over the past few years we have commenced
a survey of the Galactic 
bulge in the near-IR.
We image the clusters using SOFI at the ESO/NTT to obtain accurate J,H,K photometry of 
globular clusters towards the bulge down to the base 
of the Red Giant Branch (RGB), 
to properly define their reddening, distance and photometric metallicity 
\citep{val04a,val04b,val05}. 
We use NIRSPEC, a high throughput infrared (IR) echelle
spectrograph at the Keck Observatory \citep{ml98} to measure their 
detailed chemical abundances.
The near IR spectral range is indeed
the most suitable to study obscured 
stellar populations, like the Galactic bulge and center.

H--band (1.5--1.8 $\mu$m) spectra
of cool giants   
contain several atomic and molecular lines for detailed abundance analysis of Fe, 
C, O and other $\alpha $-elements. 
The abundance and abundance pattern distributions in the {\it cluster} and {\it field}
populations are indeed important in constraining the history of bulge formation and 
chemical enrichment \citep{mw97}.
We have used this method to derive abundances for
six bulge globular clusters: the results for NGC~6553 
and Liller~1 are given in \citet{ori02}, abundances for 
Terzan~4 and Terzan~5 are reported in \citet{ori04}, while 
those for NGC~6342 and NGC~6528 are given in \citet{ori05}.
We find $\alpha$-enhancement at a level of a factor between 2 and 3 over 
the whole range of metallicity spanned by the clusters in our survey, 
from [Fe/H]$\approx$--1.6 (cf. Terzan~4) up to $\approx-0.6 < [Fe/H] < \approx$--0.1 
(cf. NGC~6342, NGC~6553, Liller~1, Terzan~5, NGC~6528).

In this paper we present the analysis of NGC~6539 and UKS~1,   
two bulge globular clusters with high reddening, possibly with intermediate metallicity between Terzan~4 
and the most metal rich ones observed so far.
NGC~6539 is a poorly studied globular cluster located in the outer Bulge region 
(l=20.8, b=+6.8).
A large dark cloud complex in the foreground
is responsible for its high reddening \citep[see][]{san87}.
Fig.~\ref{fig1} (left panel) shows the IRAS 100 $\mu m$ image of the  
$2^{\circ} \times 2^{\circ}$ field around the cluster, where 
the foreground dust complex is clearly visible.
For this cluster \citet{zw84} found [Fe/H]=--0.66$\pm$0.15 from 
integrated optical spectra.
More recently, \citet{ste04} by means of low
resolution IR spectroscopy of four giants found $\rm [Fe/H]=-0.79\pm0.09$.
UKS~1 is a very reddended globular cluster located in the inner Bulge 
(l=5.1, b=+0.8) and low onto the Galactic plane (see Fig.~\ref{fig1}, right panel).
It was discovered by \citet{mal80}  
and later on, both optical and near IR color-magnitude diagrams (CMDs) of the upper 
RGB have been published by 
\citet{min95,ori97,or01}.  
The photometric estimate of the cluster metallicity 
is strongly dependent on the reddening assumptions and controversial values of [Fe/H] 
between -1.2 and -0.3 
have been suggested, that is either a metal poor or a metal rich object.
Integrated optical spectroscopy by \citet{bic98} suggests a metallicity close to 
solar.

Both clusters lack high resolution measurements for 
accurate metallicity estimates, independent of reddening assumptions, 
and detailed abundance distributions.
In the present work homogeneous metallicities and abundance 
patterns by means of near IR photometry 
and high resolution spectroscopy are derived for both clusters.
Our observations and data reduction follow in Sect.~2.
Sect.~3 discusses our abundance and radial velocity analysis and 
in Sect.~4 we discuss our findings.

\section{Observations and Data Reduction}

\begin{figure*}
\centering
\includegraphics[width=18cm]{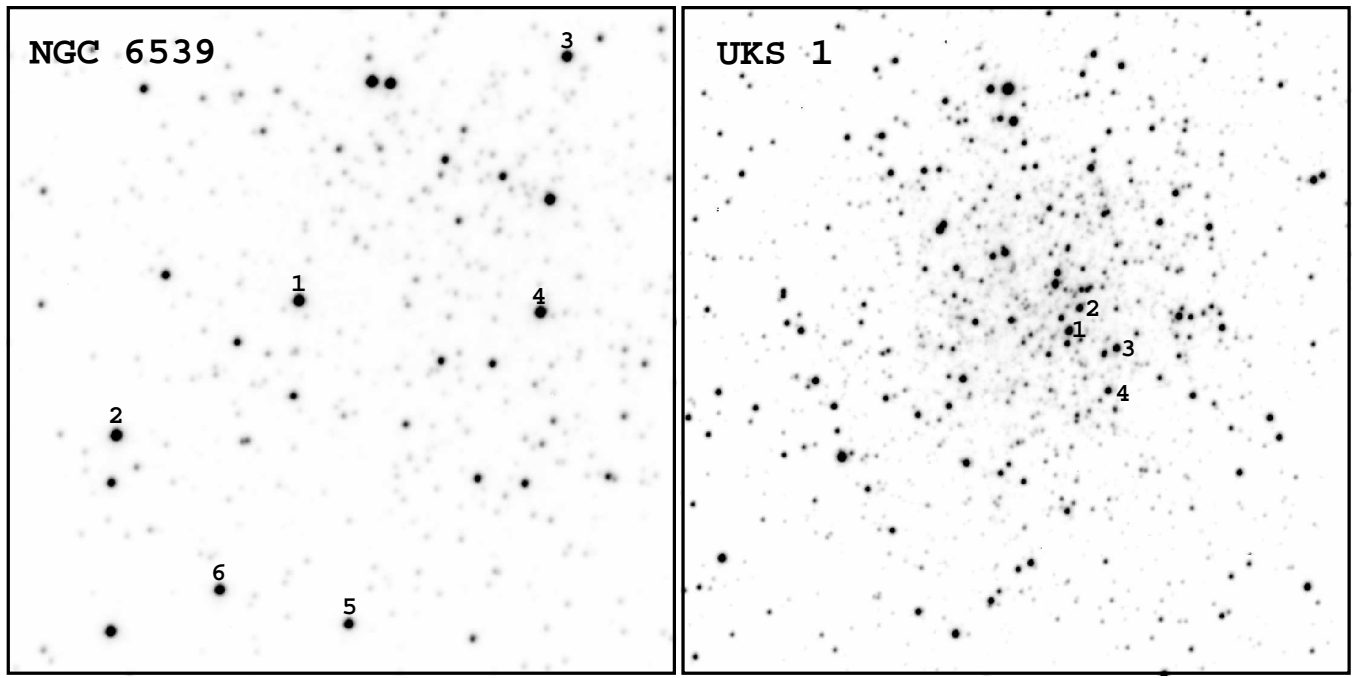}
\caption{
H band images of the core regions of NGC~6539 (left) and UKS~1 (right) 
as imaged by SOFI. 
The field of view is 90\arcsec on a side
and the image scale is
$\rm 0\farcs146~pixel^{-1}$.
The stars spectroscopically observed are numbered (cf. Table~1).
}
\label{fig2}
\end{figure*}

\subsection{Photometry}
Near IR images of 
NGC~6539 and UKS~1 were obtained 
at the European Southern
Observatory (ESO), La Silla on June 2004, with the near\--IR imager
SOFI, mounted at the ESO/NTT telescope. 
We used the large field camera
coupled with the focal elongator, yielding a pixel size of
$0 \farcs 146$ and a total field of view of $2\farcm 49 \times 2\farcm 49$.
In order to avoid saturation of the brightest RGB stars,
we secured 30 single exposures of 1.2 sec in each filter.
Every average image has been background\--subtracted
by using sky fields located several arcmin away from the cluster center,
and flat\--field corrected
using a halogen lamp alternatively switched {\it on} and {\it off}, 
following the prescription of the standard SOFI calibration setup.
Fig.~\ref{fig2}
shows the H band images of NGC~6539 (left) and UKS~1 (right).
The stars spectroscopically observed are numbered (cf. Table~\ref{tab1}). 

Standard crowded field photometry, including Point Spread Function
(PSF) modeling, was carried out on each
frame by using DAOPHOTII/ALLSTAR \citep{dao94}.
The photometric catalogues, listing the instrumental J, H and K
magnitudes, were obtained by cross\--correlating the single\--band catalogues.
By using the Second Incremental Release Point Source Catalogue of
2MASS, the instrumental magnitudes were then converted
into the 2MASS photometric system.
An overall uncertainty of ${\pm}0.05$ mag in the zero\--point
calibration in all three bands has been estimated.
The photometric catalogues have also been placed on the 2MASS astrometric system,
using a procedure developed at the Bologna Observatory
(P. Montegriffo, private communications) and successfully applied to
other clusters
\citep[see e.g.][and reference therein]{val05}
providing rms residuals of $\approx$0.2 arcsec in both R.A. and DEC.

\subsection{Spectroscopy}

High-resolution echelle spectra of six bright giants
in the core of the bulge globular
clusters NGC~6539 and four in UKS~1 (see Fig.~\ref{fig2}) have been acquired
during two observational campaigns in July 2002 and April 2004.
We used the infrared spectrograph NIRSPEC \citep{ml98}
which is at the Nasmyth focus of the Keck~II telescope.
The high resolution echelle mode, with a slit width of $0\farcs43$
(3 pixels) and a length of 24\arcsec\
and the standard NIRSPEC-5 setting, which
covers most of the 1.5--1.8 micron H-band,
has been selected.  Typical exposure times (on source)
ranged from 4 to 8 minutes. 


The raw two dimensional spectra were processed using the
REDSPEC IDL-based package written
at the UCLA IR Laboratory.
Each order has been
sky subtracted by using the pairs of spectra taken
with the object nodded along the slit, and subsequently
flat-field corrected.
Wavelength calibration has been performed using arc lamps and a second order
polynomial solution, while telluric features have been removed by
dividing by the featureless spectrum of an O star.
At the NIRSPEC resolution of R=25,000 several single
roto-vibrational OH lines and
CO bandheads can be measured to derive accurate oxygen and carbon abundances.
Other metal abundances can be derived from the atomic lines
of Fe~I, Mg~I, Si~I, Ti~I and Ca~I.
Abundance analysis is performed by using full spectral synthesis
techniques and equivalent width measurements of representative lines.

\begin{table*}
\begin{center}
\caption[]{$\rm (J-K)_0$ colors, bolometric magnitudes, heliocentric radial velocity and
equivalent widths (m\AA) of some representative lines
for the observed stars in NGC~6539 and UKS~1.}    
\label{tab1}
\begin{tabular}{lcccccccccccccc}
\hline\hline
 & & & & & & & & & & & & &\\
 & & \multicolumn{6}{c}{NGC~6539}&&&\multicolumn{4}{c}{UKS~1}\\
 & & & & & & & & & & & & &\\
\hline
 & & & & & & & & & & & & &\\
star                && \#1& \#2& \#3& \#4 &\#5&\#6&&&\#1& \#2& \#3& \#4 \\
 & & & & & & & & & & & & &\\
$\rm (J-K)_0^a$  &&1.04 &1.05 &1.00 &1.07 &0.96 &0.92 &&&1.07 &1.00 &1.05 &1.04\\
$\rm M_{bol}^b$ &&--3.0 &--2.9 &--2.6 &--2.9 &--2.3 &--2.6 &&&--3.0 &--2.7 &--2.9 &--2.6\\
$v_r$ [km~s$^{-1}$] &&+43 &+27 &+33 &+24 &+27 &--135$^c$ &&&+45 &+67 &+58 &+187$^c$\\
Ca~$\lambda $1.61508 &&146 &146  &149 & 144 &126 &166 &&&156 &121 &143 &267\\
Fe~$\lambda $1.61532 &&174 &178  &188 & 170 &188 &216 &&&164 &168 &166 &269\\
Fe~$\lambda $1.55317 &&153 &164  &165 & 141 &171 &194 &&&149 &157 &154 &228\\
Mg~$\lambda $1.57658 &&401 &400  &401 & 388 &404 &426 &&&372 &386 &385 &453\\
Si~$\lambda $1.58884 &&470 &479  &478 & 489 &453 &478 &&&432 &416 &422 &527\\
OH~$\lambda $1.55688 &&264 &270  &264 & 302 &200 &228 &&&257 &181 &243 &329\\
OH~$\lambda $1.55721 &&261 &268  &263 & 306 &200 &225 &&&256 &179 &238 &330\\
Ti~$\lambda $1.55437 &&330 &328  &318 & 345 &307 &331 &&&306 &294 &292 &412\\
\hline
\end{tabular}
\end{center}
$^a$ Reddening corrected colors, 
adopting \citet{har96} 
E(B-V)=1.00 for NGC~6539 and E(B-V)=3.09 for UKS~1.\\
$^b$ Bolometric magnitudes adopting \citet{har96} distance moduli of (m-M)$_0$=14.36 
for NGC~6539 and (m-M)=14.39 for UKS~1 and bolometric correction from \citet{mon98}.\\
$^c$ Unlikely cluster members.
\end{table*}

\subsubsection{Abundance Analysis}
\label{analysis}

We computed suitable synthetic spectra
of giant stars by varying the stellar parameters and the
element abundances using an updated
version of the code described in \citet{ori93}.
The main characteristics of the code have been widely discussed in our previous 
papers 
\citep{ori02,ori04,ori05} and they will not be repeated here.
The code 
uses the LTE approximation and is based
on the molecular blanketed model atmospheres of
\citet{jbk80} at temperatures $\le$4000~K
and the ATLAS9 models for temperatures above 4000~K.
Recently, the NextGen model atmospheres \citep{hau99} have been also 
implemented within our code and tested. However, when compared with the older models 
minor differences (well within a few hundredths dex)    
in the resulting abundances have been found.
This is not surprising, since the major source of opacity in the H band spectra 
of cool stars is H$^-$ and small differences in the temperature structure  
of different model atmospheres have a minor impact on the overall abundance determination. 
The reference solar abundances are from
\citet{gv98}.

Photometric estimates of the stellar parameters are initially used 
as input to produce a grid of model spectra, 
allowing the abundances 
and abundance patterns to vary over a large range and the stellar parameters 
around the photometric values.  
The model which better reproduces the overall observed spectrum and 
the equivalent widths of selected lines is chosen as the best fit model.
We measure equivalent widths in the observed spectrum (see Table~\ref{tab1}), 
in the best fit model and in four additional models which are, respectively, 
$\pm$0.1 and $\pm$0.2 dex away from the best-fitting.
This approach gives us the random errors in the inferred abundances
listed in Table~\ref{tab2}.

\begin{figure*}
\centering
\includegraphics[width=18cm]{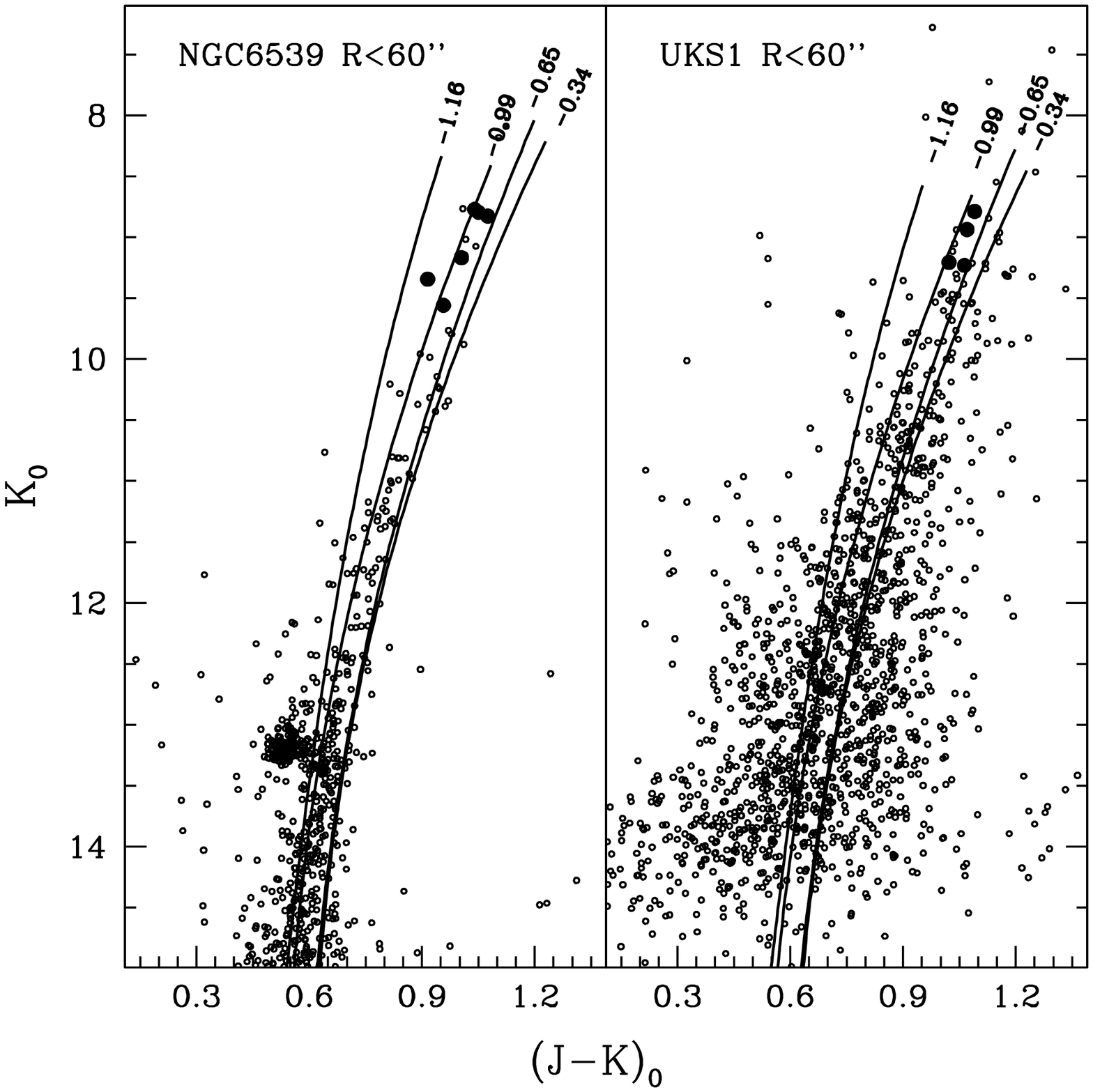}
\caption{
$\rm K_0,(J-K)_0$ de-reddened CMDs of NGC~6539 (left) 
and UKS~1 (right) in the central 1 arcmin in radius. 
Suitable average ridge lines of reference globular clusters with different metallicities 
are reported for comparison \citep[see][and references therein]{val05}: NGC~362, [Fe/H]=-1.16; NGC~6638, [Fe/H]=-0.99;
NGC~6342 ,[Fe/H]=-0.65; NGC~6440, [Fe/H]=-0.34.
Large filled points mark the stars spectroscopically observed.
}
\label{fig3}
\end{figure*}

Stellar parameter uncertainty of $\pm$200~K in temperature (T$_{eff}$), 
$\pm$0.5 ~dex in gravity (log~g) and $\pm$0.5~km~s$^{-1}$ in 
microturbulence velocity ($\xi$), 
can introduce a further systematic $\le$0.2~dex uncertainty in the 
absolute abundances.
However, since the CO and OH molecular line profiles are very sensitive to 
effective temperature, gravity, and microturbulence variations, 
they better constrain the values of these parameters,  
significantly reducing their initial range of variation and
ensuring a good self-consistency of the overall spectral
synthesis procedure \citep{ori02,ori04,ori05}.
Indeed, our previous analysis of six globular clusters 
show that solutions with  
$\Delta $T$_{\rm eff}$$=\pm$200~K, $\Delta $log~g=$\pm$0.5~dex and
$\Delta \xi$$=\mp$0.5~km~s$^{-1}$ and corresponding 
$\pm$0.2~dex abundance variations from the best-fitting one are indeed less
statistically
significant (typically at $1\le\sigma\le3$ level only). 
Moreover, since the
stellar features under consideration show a similar trend
with variation in the stellar parameters, although with different
sensitivity, {\it relative } abundances are less
dependent on stellar parameter assumptions, 
reducing the systematic uncertainty to $<$0.1~dex.

In order to check further the statistical significance of our best-fitting
solution, we also compute synthetic spectra with 
$\Delta $T$_{\rm eff}$$=\pm$200~K, $\Delta $log~g=$\pm$0.5~dex and
$\Delta \xi$$=\mp$0.5~km~s$^{-1}$, and with corresponding simultaneous 
variations 
of $\pm$0.2~dex of the C and O abundances to reproduce the depth of the
molecular features.

We follow the strategy illustrated in \citet{ori04,ori05}.
As a figure of merit we adopt
the difference between the model and the observed spectrum (hereafter $\delta$).
In order to quantify systematic discrepancies, this parameter is indeed
more powerful than the classical $\chi ^2$ test, which is instead
equally sensitive to {\em random} and {\em systematic} scatters.
Since $\delta$ is expected to follow a Gaussian distribution,
we compute $\overline{\delta}$ and the corresponding standard deviation
for our best-fitting solution and the other models 
with the stellar parameter and abundance variations quoted above.
We then extract 10,000 random subsamples from each
{\it test model} (assuming a Gaussian distribution)
and we compute the probability $P$
that a random realization of the data-points around
a {\it test model} display a $\overline{\delta}$ that is compatible
with an ideal best-fitting model with a $\overline{\delta}$=0. 
$P\simeq 1$ indicates that the model is a good representation of the
observed spectrum.
The statistical tests are performed on portions of the spectra
mainly containing the CO bandheads and the OH lines which are the 
most sensitive to the stellar parameters.

\begin{figure*}
\centering
\includegraphics[width=18cm]{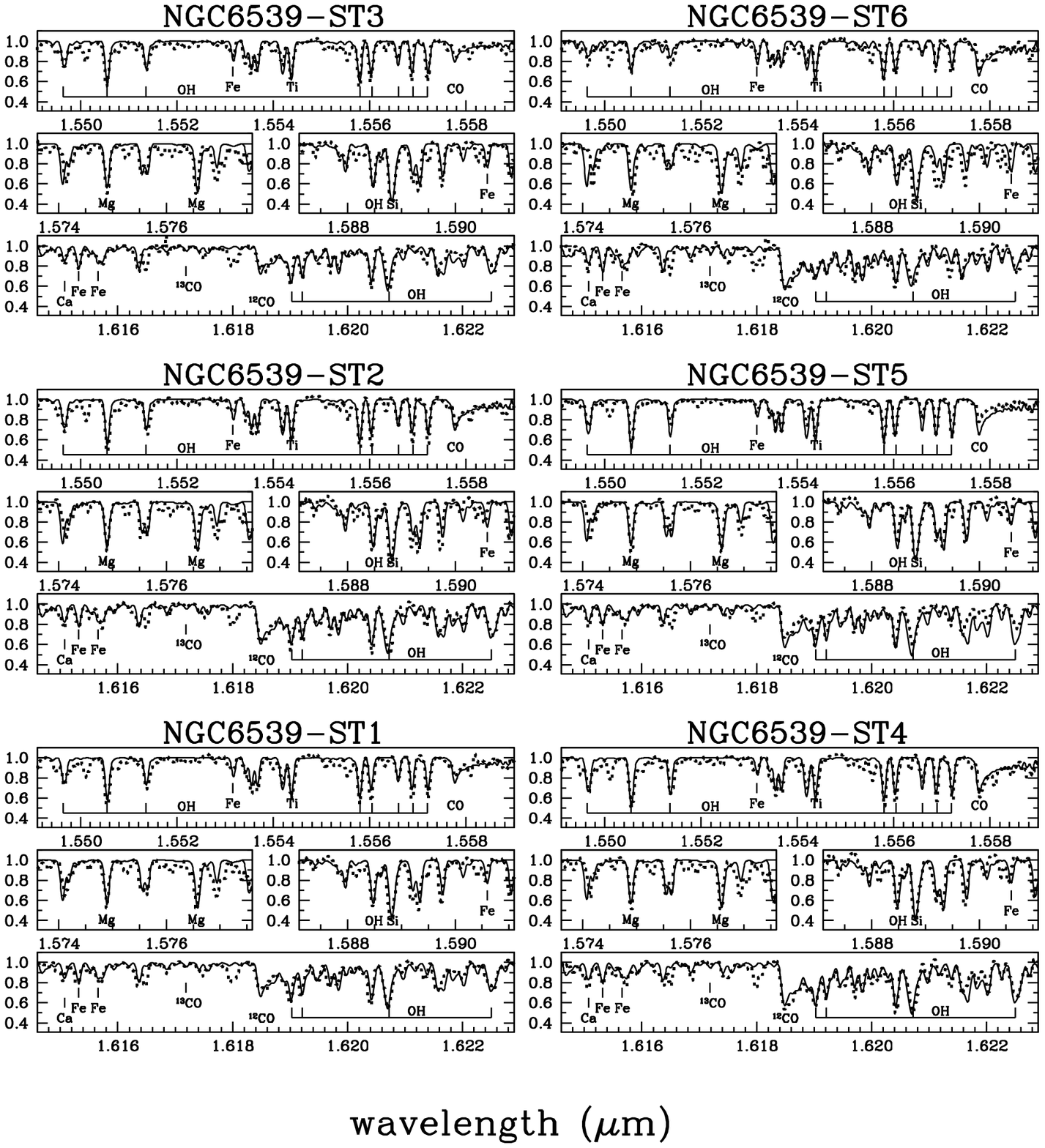}
\caption{
Selected portions of the observed echelle spectra (dotted lines) of the
six giants in NGC~6539 with our best-fitting synthetic spectrum
(solid line) superimposed. A few important molecular and atomic lines
of interest are marked.
}
\label{fig4}
\end{figure*}


\begin{figure*}
\centering
\includegraphics[width=18cm]{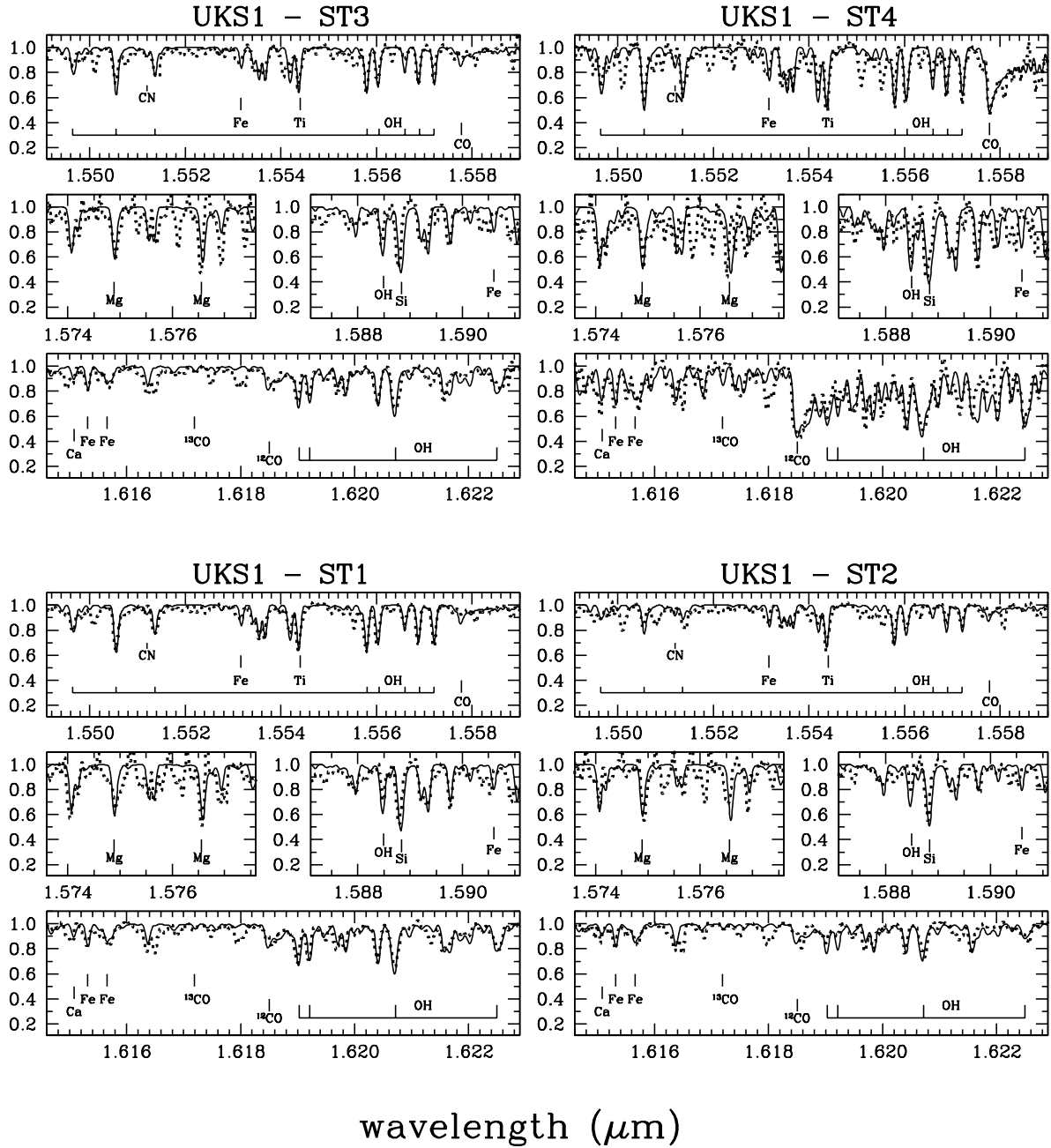}
\caption{
As in Fig.\ref{fig4} but 
for the four giants in UKS~1.
}
\label{fig5}
\end{figure*}

\begin{table*}
\begin{center}
\caption[]{Adopted stellar atmosphere parameters and abundance estimates.}
\label{tab2}
\begin{tabular}{lcccccccccccccc}
\hline\hline
 & & & & & & & & & & & & & &\\
 & & \multicolumn{6}{c}{NGC~6539}&&&\multicolumn{4}{c}{UKS~1}\\
 & & & & & & & & & & & & &\\
\hline
 & & & & & & & & & & & & &\\
star           && \#1  & \#2  & \#3  & \#4  &\#5 &\#6$^b$ &&& \#1  & \#2  & \#3  & \#4$^b$ \\
 & & & & & & & & & & & & &\\
T$_{\rm eff}$ [K] &&3800 &3800 &3800 &3600 &4000 &4000 &&&3800 &4000 &3800 &3800\\ 
log~g             &&0.5 &0.5 &0.5 &0.5 &0.5 &0.5 &&&1.0 &1.0 &1.0 &0.5\\
$\xi$ [km~s$^{-1}$]   &&2.0  &2.0  &2.0  &2.0  &2.0 &2.0 &&&2.0 &2.0  &2.0 &2.0\\
$\rm [Fe/H]$   &&--0.77&--0.74&--0.72&--0.80&--0.71&--0.53&&&--0.81&--0.78&--0.75&--0.26\\
               &&$\pm$.07&$\pm$.07&$\pm$.07&$\pm$.08&$\pm$.07&$\pm$.07&&&$\pm$.09&$\pm
$.08&$\pm$.08&$\pm$.11 \\
$\rm [O/Fe]$   &&+0.45&+0.44&+0.37&+0.40&+0.40&+0.41&&&+0.35&+0.23&+0.22&+0.30\\
               &&$\pm$.09&$\pm$.10&$\pm$.08&$\pm$.10&$\pm$.12&$\pm$.08&&&$\pm$.10&$\pm
$.09&$\pm$.09&$\pm$.12 \\
$\rm [Ca/Fe]$  &&+0.48 &+0.44&+0.40&+0.40&+0.41&+0.43&&&+0.39&+0.38&+0.35&+0.41\\
               &&$\pm$.12 &$\pm$.12 &$\pm$.12 &$\pm$.13 &$\pm$.13&$\pm$.12&&&$\pm$.17&$\pm
$.17 &$\pm$.17 &$\pm$.14 \\
$\rm [Si/Fe]$  &&+0.42&+0.44&+0.47&+0.40&+0.41&+0.41&&&+0.22&+0.28&+0.25&+0.36\\
               &&$\pm$.18 &$\pm$.18 &$\pm$.18 &$\pm$.19&$\pm$.16&$\pm$.16&&&$\pm$.18&$\pm
$.17 &$\pm$.17 &$\pm$.19 \\
$\rm [Mg/Fe]$  &&+0.46&+0.49&+0.42&+0.44&+0.46&+0.43&&&+0.29&+0.32&+0.30&+0.37\\
               &&$\pm$.18 &$\pm$.18 &$\pm$.18 &$\pm$.17 &$\pm$.16&$\pm$.16&&&$\pm$.16&$\pm
$.19 &$\pm$.15 &$\pm$.16 \\
$\rm [Ti/Fe]$  &&+0.50&+0.44&+0.33&+0.40&+0.45&+0.40&&&+0.31&+0.33&+0.25&+0.40\\
               &&$\pm$.19 &$\pm$.19 &$\pm$.19 &$\pm$.16 &$\pm$.19&$\pm$.18&&&$\pm$.19&$\pm
$.18 &$\pm$.18 &$\pm$.15 \\
$\rm [\alpha/Fe]^a$  &&+0.47&+0.45&+0.40&+0.41&+0.43&+0.42&&&+0.31&+0.33&+0.29&+0.38\\
               &&$\pm$.11 &$\pm$.11 &$\pm$.11 &$\pm$.12 &$\pm$.12&$\pm$.11&&&$\pm$.12&$\pm
$.12 &$\pm$.12 &$\pm$.14 \\
$\rm [C/Fe]$   &&--0.33&--0.26&--0.47&--0.20&-0.29&-0.27&&&--0.49&--0.52&--0.55&--0.24\\
               &&$\pm$.10 &$\pm$.10 &$\pm$.10 &$\pm$.11 &$\pm$.10&$\pm$.10&&&$\pm$.12&$\pm
$.11 &$\pm$.11 &$\pm$.13 \\
\hline
\end{tabular}
\end{center}
$^a$ $\rm [\alpha/Fe]$ is the average $\rm [<Ca,Si,Mg,Ti>/Fe]$ abundance ratio.\\
$^b$ Unlikely cluster members.
\end{table*}

\section{Results}
\label{results}

\subsection{IR Colour-Magnitude Diagrams}
We use our IR photometric catalogs to construct suitable CMDs of the two clusters.
Fig.~3 shows the $\rm K_0,(J-K)_0$ de-reddened CMDs 
in the innermost region
of the two clusters with superimposed the mean ridge lines a few 
reference clusters with different metallicities, reported at the 
adopted distance of NGC~6539 and UKS~1.
In order to do this,
for NGC~6539 we adopt E(B-V)=1.00 and (m-M)$_0$=14.48 \citep{har96}. 
The extinction of UKS~1 is huge: 
values between E(B-V)$\approx$3 and 3.4 \citep{min95,har96,ori97,or01} and
distance moduli between $\approx$14.2 \citep{har96} up to $\approx$15 \citep{min95} 
have been suggested. 
We adopt E(B-V)=3.20 and (m-M)$_0$=14.6 intermediate values.
The RGB sequence in the CMD of NGC~6539 is rather well defined down to a couple of magnitudes below the 
Horizontal Branch.
Its $\rm (J-K)_0$ colors are most likely consistent with a metallicity  
somewhat in between NGC~6638 ([Fe/H]=0.99) and NGC~6342 ([Fe/H]=-0.65).  
The CMD of UKS~1 is less deep since the cluster is much more reddened and it is more sparse 
due to a higher degree of field contamination and possibly differential reddening, 
also in the innermost core region.
Indeed, by using 2MASS photometry on an annular region between 2 and 6 arcmin 
around the cluster center, a $\rm \Delta E(J-K)$ up to $\approx$0.3~dex is present between 
the NW and NE sectors. 
The comparison of the UKS~1 RGB with the mean ridge lines of reference clusters (see Fig.~\ref{fig3}),
gives only a rough constraint of $\rm -1<[Fe/H]<-0.3$ dex.
As discussed in detail by \citet[][]{f00,val04a}, a
complete characterization of the RGB morphology, as a function of the cluster
metallicity, can be obtained
by computing a set of photometric indices, namely
{\it (i)} the colours at fixed magnitudes; {\it (ii)} the magnitudes
at constant colours and {\it (iii)} the RGB slope, 
using suitable CMDs and intrinsic ridge lines.  
The RGB slope is an interesting feature being both reddening and
distance independent. 
Given the large reddening in both clusters we use this parameter as 
an indication of the photometric metallicity. 
By using the calibration of \citet{val04a} we constrain 
[Fe/H] between --0.8 and --0.7 in both clusters.

In each CMDs we also mark the position of the stars spectroscopically observed.
They all lie in the innermost region of the clusters.
In order to obtain a photometric estimate of the stellar temperatures
and the bolometric magnitudes of these stars we adopt  
the color-temperature transformations and
bolometric corrections of \citet{mon98}, specifically
calibrated for globular cluster giants.
By using the $\rm (J-K)_0$ color, we constrain effective temperatures in the range 
3600--4000~K, and
bolometric magnitudes between -2 and -3
(see Table~\ref{tab1}), their precise values depending on the adopted 
reddening and distance.

\subsection{Chemical abundances}

Detailed abundances and abundance patterns have been obtained 
from the high resolution spectra, by combining full spectral synthesis 
analysis with equivalent width
measurements. 
We derive abundances of Fe, C, O
and $^{12}$C/$^{13}$C for the observed
giants in NGC~6539 and UKS~1.        
The abundances of additional $\alpha-$elements Ca, Si, Mg and Ti are obtained
by measuring a few major atomic lines.\\
Stellar temperatures are both estimated from the 
$\rm (J-K)_0$ colors (see Table~\ref{tab1}) and molecular lines,
gravity from theoretical evolutionary tracks,
according to the location of the stars on the RGB, 
and adopting an average microturbulence velocity of 2.0 km/s
\citep[see also][]{ori97}.
The final adopted temperatures, obtained by best-fitting the CO and in 
particular the OH molecular bands which are 
especially temperature sensitive in cool giants, are reported in
Table~\ref{tab2}.
Equivalent widths (see Table~\ref{tab1}) are computed by Gaussian fitting 
the line profiles and the overall uncertainty is $\le$10\%.

Fig.~\ref{fig4} shows
our synthetic best fit models superimposed on the
observed spectra of the six giants in NGC~6539.
Stars \#1 to \#5 have similar heliocentric radial velocities, 
(cf. Table~\ref{tab1}),
the average value 
being $\rm<v_r>=+31\pm 4~km/s$ and a velocity dispersion of $\approx$8~km/s.
Star \#6 has $\rm v_r=-135~km/s$, so it is unlikely to be a member of the cluster. 
From our overall spectral analysis (cf. Table~\ref{tab2}) 
we find the average cluster $\rm [Fe/H]=-0.76\pm0.03$, $\rm [O/Fe]=0.43\pm 0.02$ and 
$\rm [\alpha/Fe]=0.44\pm0.02$
Star \#6 has a slightly higher metal content but similar $\alpha$-enhancement.
We also measure an average carbon depletion ([C/Fe]=--0.30$\pm0.09$~dex)
and low $\rm ^{12}C/^{13}C\approx 4.4\pm0.7$.
Fig.~\ref{fig5} shows
our synthetic best-fitting superimposed on the
observed spectra of the four giants in UKS~1.
Stars \#1 to \#3 are likely cluster members 
(cf. Table~\ref{tab1}),
with an average $\rm<v_r>=+57\pm 6~km/s$
and a velocity dispersion of $\approx$11~km/s.
Star 4 has a significantly higher radial velocity, $\rm v_r=+187$~km/s.
For this cluster our abundance analysis 
(cf. Table~\ref{tab2}),
give an average $\rm [Fe/H]=-0.78\pm0.03$,
$\rm [O/Fe]=+0.27\pm0.07$ and an overall average $\rm [\alpha/Fe]=0.31\pm0.02$.
We also measure an average carbon depletion ([C/Fe]=--0.52$\pm0.03$~dex),
a low $^{12}$C/$^{13}$C$\approx4.7\pm0.8$. 
Star \#4 has a significantly higher iron abundance and 
slightly higher [$\alpha$/Fe] and [C/Fe] abundance ratios.

\begin{figure}
\centering
\includegraphics[width=8.7cm]{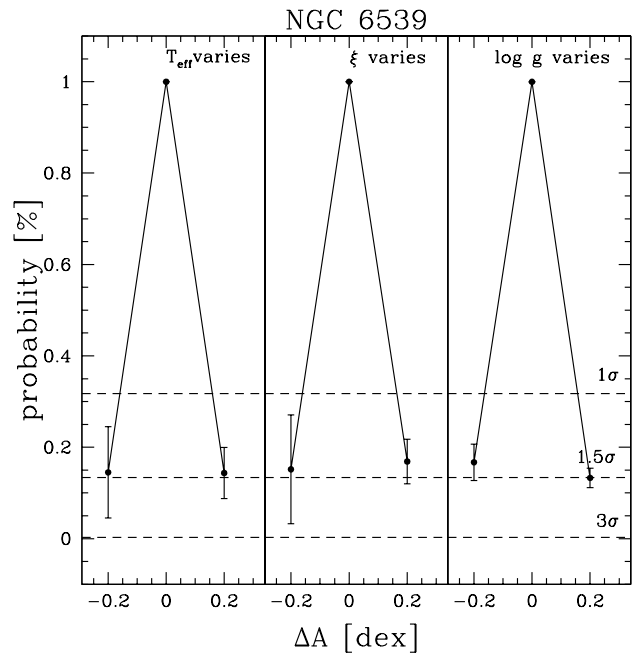}
\caption{
Average probability of a random realization
of our best-fitting solution and the test models with varying temperature by 
$\Delta T_{eff}$ of $\pm$200K (left panels), 
gravity by $\Delta log~g$ of $\pm$0.5 dex (middle panels), 
and microturbulence by $\Delta \xi$ of $\mp$0.5 Km s$^{-1}$ (right panels),  
with respect to the best-fitting (see Sect.~\ref{results}) for the five probable 
members of NGC~6539.
}
\label{fig6}
\end{figure}

\begin{figure}
\centering
\includegraphics[width=8.7cm]{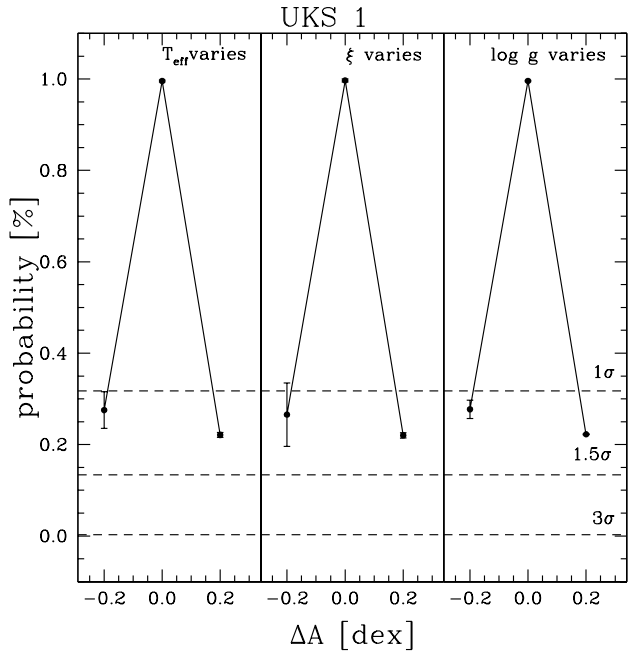}
\caption{
As in Fig.~\ref{fig6}, but for
three giants most likely members of UKS~1 (see also Sect.~\ref{results}).
}
\label{fig7}
\end{figure}

As shown in Figs.~\ref{fig6} and \ref{fig7} our best-fitting solutions for 
both clusters  
have an average probability $\-P>$0.99 to be statistically 
representative of the observed spectra. 
The other {\it test models} with slightly different assumptions for stellar temperatures, 
gravity, microturbulence and
photospheric abundances (see Sect.~\ref{analysis}) are only significant at $1<\sigma<1.5$ level. 

\section{Discussion and Conclusions}

We report near-IR photometry and high resolution spectroscopy
for NGC~6539 and UKS~1. 
Both the RGB slope and abundance analysis of likely cluster members are consistent with them 
having similar iron abundance of $\approx$1/6 solar.
Our iron abundance of NGC~6539 is fully consistent with the value 
found by \citet{ste04}.  
The significantly higher value suggested 
by \citet{bic98} for UKS~1 and obtained from integrated spectroscopy, can be 
easily explained in terms of field contamination. 
Indeed, our star \#4, which is not a cluster member despite its position towards 
the cluster center and its luminosity typical of a RGB star near the tip, 
has a significantly higher metallicity, close to the value of \citet{bic98}.  
According to our IR CMDS (cf. Fig.\ref{fig3})
a [Fe/H]=--0.78 for UKS~1 is consistent with $\rm 1.52<E(J-K)<1.54$, e.g. 
$\rm 3.1<E(B-V)<3.15$, using a E(J--K)/E(B--V)=0.49 reddening law \citep{sv}.
Both clusters show enhancement of $\alpha$-elements, confirming the scenario 
that the Bulge undergoes a rapid chemical enrichment from type~II SNe ejecta.

The low $\rm^{12}C/^{13}C$ abundance ratios measured in NGC~6539 and UKS~1 are similar to 
those found in our analysis of the other bulge clusters 
\citep[see also][]{she03}
as well as to those measured in the halo  
globular clusters 
\citep{she96,gra00,vws02,smi02,ori03}.
They can be explained by additional mixing mechanisms due to 
{\it cool bottom processing} in the stellar interiors during the evolution along
the RGB \citep[see e.g.][]{cha95,dw96,csb98,bs99}, and occurring 
over the entire metallicity range between one hundredth solar and solar.

Older measurements of radial velocities in NGC~6539, based on low resolution spectra
\citep{zw84,hes86} suggest somewhat different values ($\rm v_r$=--35 and --51~km/s, respectively)  
than those obtained here (see Table~1). However, since both the resolution and the quality of 
their spectra are low, as stated by the authors themselves,  
our new estimate of $\rm v_r$=+31 based on high resolution spectra should be more accurate.  

Finally, we would like to note that stars \#6 in NGC~6539 and star \#4 in UKS~1, 
altough unlikely cluster members, have radial velocities and abundance patterns consistent 
with the field Bulge population \citep{ro05}.

\section*{Acknowledgments}

LO, EV and FRF acknowledge the financial support by 
the Ministero dell'Istru\-zio\-ne, Universit\`a e Ricerca (MIUR).\\
RMR acknowledges support from grant number AST-0098739,
from the National Science Foundation.
The authors are grateful to the NTT staff
of the ESO Observatory.
They also acknowledge the Keck Observatory and 
the NIRSPEC team.
The authors wish to recognize and acknowledge the very significant cultural
role and reverence that the summit of Mauna Kea has always had within
the indigenous Hawaiian community.
We are most fortunate to have the opportunity to conduct observations 
from this mountain.
This publication makes use of data products from the Two Micron All Sky Survey,
which is a joint project of the University of Massachusetts and Infrared
Processing and Analysis Center/California Institute of Technology, founded by
the National Aeronautics and Space Administration and the National Science
Foundation.

\label{lastpage}

\end{document}